\documentclass[12pt,preprint]{aastex}
\newcommand\msol{M$_{\odot}$}
\newcommand\ds{dSph}
\newcommand\di{dIrr}

\newcommand\kms{km~s$^{-1}$}
\renewcommand\deg{$^{\circ}$}
\newcommand\hi{H{\sc i}}

\begin{document}

\title{Neutral Hydrogen Clouds near Early-Type Dwarf Galaxies of the Local
Group}
\shorttitle{H{\sc i} near Local Group dwarfs}

\author{Antoine Bouchard}
\affil{Research~School~of~Astronomy~\&~Astrophysics,
The~Australian~National~University, Mount~Stromlo~Observatory, Cotter~Road,
Weston~Creek, ACT~2611, Australia\\and\\ Australia~Telescope~National~Facility,
CSIRO, PO~Box~76, Epping, NSW~1710, Australia}
\email{bouchard@mso.anu.edu.au}

\author{Claude Carignan}
\affil{D\'epartement~de~physique~and~Observatoire~du~mont~M\'egantic, 
Universit\'e~de~Montr\'eal, C.P.~6128, Succ.~Centre-ville, Montr\'eal, Qu\'ebec, Canada H3C~3J7} 
\email{claude.carignan@umontreal.ca}

\author{Lister Staveley-Smith}
\affil{Australia~Telescope~National~Facility, 
CSIRO, PO~Box~76, Epping, NSW~1710, Australia}
\email{lister.staveley-smith@csiro.au}

\begin{abstract}
Parkes\footnote{The Parkes telescope is part of the Australia Telescope
which is funded by the Commonwealth of Australia for operation as a
National Facility managed by CSIRO.} neutral hydrogen 21 cm line (\hi{})
observations of the surroundings of 9 early-type Local Group dwarfs are
presented.  We detected numerous \hi{} clouds in the general direction of
those dwarfs and these clouds are often offset from the optical center of
the galaxies. Although all the observed dwarfs, except Antlia, occupy
phase-space regions where the High-Velocity Cloud (HVC) density is well
above average, the measured offsets are smaller than one would expect from
a fully random cloud distribution.  Possible association is detected for 11
of the 16 investigated clouds, while, for two galaxies, Sextans and Leo I,
no \hi{} was detected.  The galaxies where \hi{} clouds were found not to
coincide with the optical, yet have a significant probability of being
associated are: Sculptor dSph, Tucanna, LGS3, Cetus, and Fornax. If the
clouds are indeed associated, these galaxies have \hi{} masses 
of $M_{\rm HI} = 2\times10^5~M_{\odot}$, $M_{\rm HI} = 2\times10^6~M_{\odot}$,
$M_{\rm HI} = 7\times10^5~M_{\odot}$, $M_{\rm HI} = 7\times10^5~M_{\odot}$, and
$M_{\rm HI} = 1\times10^5~M_{\odot}$, respectively. However, neither ram
pressure nor tidal stripping can easily explain the offsets. In some cases,
large offsets are found where ram pressure should be the least effective.

\end{abstract}

\section{Introduction}

Early type dwarf spheroidal (\ds) and mixed type dwarf irregular/dwarf
spheroidal (\di/\ds) galaxies of the Local Group (LG) are torn apart by the
presence of all the other members of the group. Their shallow potential well
makes them good targets for many kind of disruptions, from tidal forces to ram
pressure due to the intergalactic medium (IGM). In general, \ds{} are found near
larger galaxies \citep{mateo98}. More precisely, the Milky Way has 7 dSph closer
than 200 kpc, which is more or less the size of its extended dark matter halo
\citep{zaritsky99}, where one would expect that disruption processes are
maximized.

\citet[hereafter BR]{blitz2000} compiled evidence that 10 of the 21 dSph and
dIrr/dSph galaxies of the LG contain important amounts of neutral hydrogen (\hi)
situated in reservoirs that can be as far as 10 kpc away from the optical core
of the galaxy. The offset was mainly attributed to ram pressure but the authors
also consider a tidal stripping mechanism. This followed the discovery of \hi{}
clouds near Sculptor \citep{carignan98b,bouchard2003}, Phoenix \citep{julie99},
SagDIG \citep{young97a}.  These detections have implications on the formation
and evolution of the LG. BR argued that most likely ``all of the LG dwarf
galaxies have had loosely bound \hi{} envelopes'' and that some stripping
mechanism removed the gas for galaxies closer than 250 kpc from the Milky Way
(MW) or M31 consequently stopping any ongoing star formation
\citep[see][]{grebel98,grebel2001}.  This scenario may be consistent with
numerical simulations of galactic formation in a cold dark matter cosmogony that
predict many more galactic halos than what is currently observed
\citep{navarro96}. A large number of substructures may also reside in the halo
of the MW but most of them must have failed to form stars thus avoiding
detection \citep{moore99}. According to BR, these halo substructures may harbor
the \hi{} clouds known as High Velocity Clouds (HVCs).

\citet{murali2000} noted that ram pressure stripping might not be as important
as what one might think. The author revised the density of the gas present in
the galactic halo to a value of $n_h < 10^{-5}$ cm$^{-3}$ at a distance of 50
kpc, a factor of 10 lower than previously estimated. This leads to the
conclusion that tidal forces might be dominating over ram pressure, at least in
the case of the Magellanic stream. Though the implication for dSphs is not that
clear, the same kind of influence could be present.  Similarly,
\citet{dercole99} showed that the gas removal occurring in a Blue Compact Dwarf
galaxy by its starburst phase --- resulting in a galactic wind --- may also not
be as devastating as previously thought. It is known that the wind from massive
stars, combined with the effect of type II supernovae (SNe) release an energy in
the interstellar medium (ISM) that is greater than the binding energy of the gas
to the galaxy and could therefore remove a large quantity of it
\citep{larson74}. But the simulations showed that, after a time of about 100 Myr
after the starburst, a large fraction of the ISM falls back on the galaxy. Of
course, these simulations consider a single isolated galaxy, free of tidal
interactions with its neighbors. In the case of dSph those tidal forces could
hold the gas and prevent it from falling back towards the galaxy.

This is supported by star formation history (SFH) analysis of some dSph in the
LG.  The stellar population of these dwarfs seems more complicated than a single
early starburst phase \citep{grebel98,hurley99,martinez99,hensler2004}.  Since
there is some star formation going on in these galaxies and since star formation
does not efficiently destroy the ISM \citep{matzner2000} it is therefore
possible that clouds of hydrogen are still present in the vicinity of dwarfs. As
suggested by BR, the large number of HVCs \citep{putman2002} could be
attributed, in part, to these \hi{} clouds falling onto the MW.

In section 2, \hi{} observations of several \ds{} and mixed type \di/\ds{} in
the LG are presented.  Results and comments on each detection are presented in
section 3. Finally, a general discussion about \hi{} gas removal can be found in
section 4, along with comments on High Velocity Clouds.

\section{Observations}
\subsection{The sample}

The main goal of this paper is to look for \hi{} clouds in or near \ds{} and
\di/\ds{} galaxies in the LG. BR have already begun this work, suggesting that
many of these galaxies should be now considered as gas rich systems. They found
\hi{} clouds, using the Leiden-Dwingeloo \hi{} survey (LDS) near And III, And V,
Leo I and Sextans and they confirmed the \hi{} detections of DDO210 and Pegasus
\citep{lo93}, LGS3 \citep{hulsbosch88}, Phoenix \citep{carignan91}, Sculptor
\citep{carignan98b} and Tucana \citep{oosterloo96}.
The LDS covers the whole northern sky and reaches a declination of -30\deg.
This makes a perfect complement to the \hi{} Parkes All Sky Survey (HIPASS)
which covers all the southern sky up to a declination of +25\deg, with
the northern extension.

\begin{table*}[tbp]
\caption{\ds{} and transitional \di/\ds{} galaxies of the local group. \citep[and reference therein]{mateo98}}\label{target}
\begin{minipage}{\linewidth} 
\renewcommand{\thefootnote}{\thempfootnote}
\renewcommand{\footnoterule}{\rule{1cm}{0cm}\vspace{-0.2cm}}
\begin{tabular}{l c c c c c c c c}
\hline
\hline
Name & $\alpha_{2000}$ & $\delta_{2000}$ & $l$ & $b$ & Type & Distance & V$_{\odot}^{opt}$ & V$_{\odot}^{radio}$\\
 & & & & & & (kpc) & (\kms) & (\kms) \\
\hline
\multicolumn{9}{c}{\sl Current sample}\\
\hline
Cetus\footnote{See \citet{whiting99}} & 00 26 11 & -11 02.6 & 101.5 & -72.9 & \ds & 775$\pm$50 & - & -\\
Sculptor	& 01 00 09 & -33 42.5 & 287.5 & -83.2 & \ds	   & 79$\pm$4   & 108$\pm$3 & 104$\pm$1\footnote{Given in \citet{bouchard2003}}\\
LGS 3		& 01 03 53 & +21 53.1 & 126.8 & -40.9 & \di/\ds    & 810$\pm$60 & -282$\pm$4 & -272$\pm$6\\
Phoenix		& 01 51 06 & -44 26.7 & 272.2 & -68.9 & \di/\ds    & 445$\pm$30 & -52$\pm$6\footnote{Given in \citet{gallart2001}}& -23$\pm$2\\
Fornax		& 02 39 59 & -34 27.0 & 237.1 & -65.7 & \ds        & 138$\pm$8  & 53$\pm$3 & -\\
Carina		& 06 41 37 & -50 58.0 & 260.1 & -22.2 & \ds        & 101$\pm$5  & 224$\pm$3 & -\\
Antlia		& 10 04 04 & -27 19.8 & 263.1 & +22.3 & \di/\ds    & 1235$\pm$65 & 351$\pm$15\footnote{Given in \citet{tolstoy2000}} & 361$\pm$2 \\
Leo I		& 10 08 27 & +12 18.5 & 226.0 & +49.1 & \ds        & 250$\pm$30 & 286$\pm$2 & -\\
Sextans		& 10 13 03 & -01 36.9 & 243.5 & +42.3 & \ds        & 86$\pm$4   & 227$\pm$3 & -\\
Aquarius\footnote{Also called DDO210} & 20 46 46 & -12 51.0 & 34.0  & -31.3 & \di/\ds    & 800$\pm$250 & -& -137$\pm$3\\
Tucana		& 22 41 50 & -64 25.2 & 322.9 & -47.4 & \ds        & 880$\pm$40 & - & -\\
Pegasus		& 23 28 34 & +14 44.8 &  94.8 & -43.5 & \di/\ds    & 955$\pm$50 & - & -182$\pm$2\\
\hline
\multicolumn{9}{c}{\sl Unobserved objects}\\
\hline
And III		& 00 35 17 & +36 30.5 & 119.3 & -26.2 & dSph       & 760$\pm$40 & -351$\pm$9\footnote{Given in \citet{cote2000}} & -341$\pm$6\footnote{Given in \citet{blitz2000}}\\
NGC 185		& 00 38 58 & +48 20.2 & 120.8 & -14.5 & dSph/dE3p  & 620$\pm$25 & -210$\pm$7 & -204$\pm$4\\
NGC 205		& 00 40 22 & +41 41.4 & 120.7 & -21.1 & E5p/\ds-N  & 815$\pm$35 & -242$\pm$3 & -229$\pm$5\\
And I		& 00 45 43 & +38 00.4 & 121.7 & -24.9 & dSph	   & 805$\pm$40 & - & - \\
And II		& 01 16 27 & +33 25.7 & 128.9 & -29.2 & \ds	   & 525$\pm$110 & - & -\\
Leo II		& 11 12 29 & +22 09.2 & 220.2 & +67.2 & \ds        & 205$\pm$12 & 76$\pm$2 & -\\
Ursa Minor	& 15 09 11 & +67 12.9 & 105.0 & +44.8 & \ds        & 66$\pm$3   & -248$\pm$2 & -\\
Draco		& 17 20 19 & +57 54.8 & 86.4  & +34.7 & \ds        & 82$\pm$6   & -293$\pm$2 & -\\
Sagittarius	& 18 55 03 & -30 28.7 & 5.6   & -14.1 & \ds-N      & 24$\pm$2   & 140$\pm$5 & -\\
\hline
\end{tabular}
\end{minipage}
\end{table*}

We first used HIPASS to identify any possible \hi{} clouds that may be
associated with a LG dwarf. In the cases where the \hi{} emission was not
directly overlapping with the optical emission we have made high
resolution, high sensitivity follow-up observations using the Parkes
Narrowband system. These latter observations were used to confirm the
HIPASS detections and analyse their velocity structure.  The list of the
observed galaxies can be found in Table \ref{target} which lists the name
of the galaxies, their equatorial and galactic coordinates, their
morphological classification, radial distance, heliocentric optical radial
velocity (V$^{opt}_{\odot}$) and heliocentric \hi{} radial velocity using
the radio definition (V$^{HI}_{\odot}$). 

\subsection{HIPASS Observations}
The \hi{} Parkes All-Sky Survey (HIPASS) is a large blind \hi{} survey covering
the whole southern sky with a beam of 15.5\arcmin \citep{barnes2001} and a
velocity coverage of -1200 to +12700 \kms{} in 1024 channels 13.2 \kms{} wide
\citep{staveley2000}. This was done with the Parkes radiotelescope, a 64 m
antenna, equipped with a 13 beam receiver. The sky is observed by scanning
regions of 8\deg{} along the sky at a rate of 1\deg/min. Each point in the sky
was observed 5 times for a total integration time of 460 s beam$^{-1}$. This
gives an rms noise of about 13 mJy beam$^{-1}$.

\subsection{Parkes Narrowband Observations}
The Parkes narrowband system is a high spectral resolution alternative to
the normal multibeam setting, used in HIPASS observations.
It has a total bandwidth of 8 MHz in 2048 channels and 2 polarizations (XX
and YY) and only the inner 7 beams were used. 

The data were reduced with the LIVEDATA reduction package distributed with the
AIPS++ software. The cube was formed using $4\arcmin \times 4\arcmin \times$
0.82 \kms{} pixels. Frequency switching mode and heliocentric velocity frame
were used.  The data have a resolution (FWHM) of 14.1$\arcmin$ spatially and
1.12 \kms{} on the spectral axis. The RMS noise is 20 mJy/beam. 

The frequency-switched data were affected by low-level ripples with a period of
a few MHz.  These were removed using two methods.  Firstly, nearby regions with
no \hi{} emission were identified.  Spectra from these regions were used as a
template baseline for regions with \hi{} emission. Secondly, parabolic baselines
were fit to the residual data.

In the case of Fornax, special attention was required in order to remove the MW
emission from the data cube.  We computed the
average spectra over the whole region covered by our data cube and subtracted it
at every pixel position. This gives us a cube where the mean intensity is zero
for every velocity channel.  To estimate the \hi{} mass of a certain cloud we
need to resubtract the mean intensity of the background at each channel where
emission is found.

\section{Results}
\begin{figure}
\begin{center}
\includegraphics[angle=0, width=0.90\textwidth]{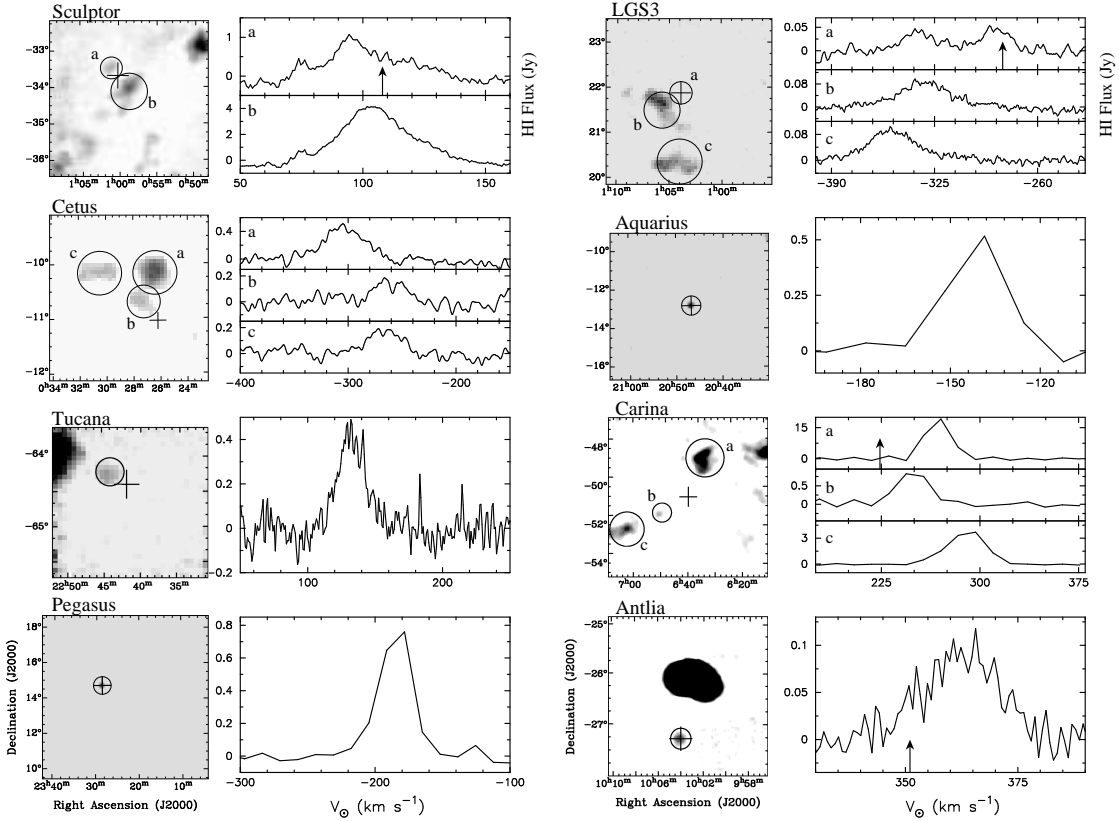}
\caption{Velocity integrated maps for a large region surrounding \ds{} and
\di/\ds{} galaxies and associated spectra. The crosses mark the position of the optical center of the
dwarfs and each investigated cloud has been circled. If more than one cloud
can be found near the center of the galaxy, a letter is given for easy
reference. On the right side of each moment map, the integrated spectra of
circled clouds are shown. An arrow marks the position of the optical
velocity in the spectra when it is known.}\label{mometspectres}
\end{center}
\end{figure}

The results of our systematic \hi{} search around \ds{} and \di{}/\ds{} galaxies
are presented in Figure \ref{mometspectres} where the \hi{} distribution maps
and the spectra for the located clouds are shown. Table \ref{result} contains
quantitative information on each cloud.

\subsection{Sculptor}
Two clouds of \hi{} were previously detected in the field of Sculptor and are
believed to be bound to the galaxy \citep{carignan98b}.  \citet{bouchard2003}
showed that 88\% of this \hi{} is contained, in projection, inside the optical
radius of the galaxy.  The very close agreement between the optical and the
\hi{} radial velocity also supports the hypothesis that the clouds are
physically linked to the dwarf.

The Magellanic Stream lies in the same general direction as Sculptor. In that
region, the Stream has a radial velocity of $\sim$0 \kms{} with a velocity
dispersion of $\sim$15 \kms \citep{bruens2005}.  It is therefore unlikely that
the Sculptor clouds belong to that structure. The other most plausible
alternative to physical association --- the clouds being HVCs --- will be
discussed in section 4.3.

\subsection{Cetus}
Recently discovered \citep{whiting99}, the Cetus dwarf is one of the latest
addition to the LG. It is classified as a \ds{} system because of its
similarities to all the other known dSph, Tucana in particular. The 2 galaxies
have roughly the same absolute magnitude and they are, strangely for \ds,
isolated systems.

The star formation history analysis reveals no obvious recent star formation in
this galaxy \citep{sarajedini2002,harbeck2004}. But as the other dSphs of the
LG, one can assume that it is not a single, short burst that shaped the galaxy.
It is affected by the second-parameter effect, probably linked with age, and
makes Cetus younger than what was expected.  It was noted by \citet{whiting99}
that because this galaxy does not seem to have any ongoing star formation, it
would be unlikely to find \hi{} within the galaxy. They have also reported a
tidal radius of $\sim$5\arcmin{} and the clouds are situated well outside of it.
The three clouds are situated less than 1.5\deg{} (a projected distance of 20
kpc) away from the dwarf and cloud b is elongated in the direction the galaxy's
optical center. 

Alone, the above is not sufficient evidence to prove that the clouds belong to
the Cetus system. No definite statement can and should be done before optical
velocities are obtained for the stars in Cetus.  When considering the measured
velocities of the \hi{} clouds, the optical velocity of Cetus should be found in
the range of $-280\pm40$ \kms{}. If this is not the case, there is little chance
that the galaxy and the clouds are related. This velocity is similar to Cetus' 2
closest LG neighbours, WLM and IC1613.

\subsection{Tucana}
\citet{oosterloo96} were the first to report a reservoir of
1.5$\times~10^{6}~[d/880$ kpc$]^2$~\msol{} of \hi{} near Tucana but they
believed it associated with the Magellanic Stream --- this would make the \hi{}
mass much lower, $\sim6\times10^{3}~[d/55$ kpc$]^2$ \msol. They were using the
Australia Telescope Compact Array (ATCA) with the 375m configuration, which
will underestimate the flux on scales larger than $\sim$10\arcmin{}.
Our observation of the Tucana dwarf, using a single dish
telescope, is therefore more precise and gives a mass of M$_{\rm HI}$ =
1.79$\pm$0.15 $\times\ 10^{6}~[d/880$ kpc$]^2$ \msol. We can conclude that 20\%
of the \hi{} resides in structure inaccessible to the array.

Apart from the artifacts at the edge of the data cube, most prominent in
the north-east corner (see Figure \ref{mometspectres}), the region around
Tucana is mostly free of \hi{} features. In this paper, we consider the
detected cloud to be associated with Tucana as we would expect the
Magellanic stream to be a larger or non-isolated low surface brightness
feature.

\subsection{LGS 3}
Previous \hi{} observations have already shown that LGS~3 is a gas rich galaxy
\citep{lo93}. We confirm that there is an \hi{} cloud at the same position and
velocity than the optical galaxy.  Figure \ref{mometspectres} shows 2 other
clouds that might be associated with LGS~3. These clouds were not discussed by
\citet{lo93} and do not appear in their spectra but were presented earlier by
\citet{hulsbosch88}.  They are spatially close to the galaxy but at sightly
different velocities. The cloud centered on LGS~3 (Cloud a) shows a blueshifted
feature that is probably an overlap from Cloud b.

A careful analysis of the \hi{} emission map shows that Clouds b and c are
linked to each other by faint emission. The shape of the clouds also
suggests that they are in physical association with the galaxy.

\subsection{Carina}
Despite previous attempts \citep[e.g.][]{mould90}, no \hi{} has ever been
reported in the vicinity of Carina.  We used the HIPASS survey to search for
\hi{} near the dwarf. Higher resolution data is not available for this galaxy.
Nevertheless, we found 3 clouds in the vicinity of Carina though they are
situated far from the optical center of the dwarf.

\citet{majewski2000} have observed the Carina dwarf in search of a tidal break
in the distribution of the stars. They found a departure from a King profile at
a radial distance of 20\arcmin{} but the stellar density profile continues to
fall off more gradually up to a radius of 80\arcmin. According to the model by
\citet{johnston99}, objects beyond this break are unbound to the galaxy and
should be considered as extratidal.  Unfortunately, they were unable to complete
the spectroscopic study needed to determine any rotation-like motion that would
be expected from extratidal debris.

The three clouds circled in Figure \ref{mometspectres} are the three closest
candidates for \hi{} being related to the \ds. One one hand, two of these clouds
(a and b) are situated near the edge of the 80\arcmin{} optical radius of the
galaxy and have velocities very close to the optical velocity of Carina
(specially for Cloud b). On the other hand, there is a large \hi{} complex
projected near the galaxy that can be partly seen in the \hi{} distribution map
and the velocity of Cloud c is drastically different to that of the dwarf.
Moreover there is no \hi{} that can be found {\it within} the 80\arcmin{} radius
of the galaxy. It is therefore doubtful that the gas and the dwarf are
physically associated.

\subsection{Phoenix}

The suggestion that \hi{} might be associated with Phoenix goes back nearly 15
years \citep{carignan91}. It was shown that the detected gas was well
separated from the MW and from the Magellanic Stream gas. This was later
confirmed, at higher resolution, by \citet{julie99}.  Despite the spatial
coincidence, it was stellar spectroscopy \citep{gallart2001} that confirmed the
likelihood of the association (V$^{opt}_{\odot} = -52\pm6$ km s$^{-1}$,
V$^{HI}_{\odot} = -23$ km s$^{-1}$).

We have not attempted to confirm the detection of \hi{} in Phoenix with the
HIPASS data because the velocity resolution is too poor, making it difficult
to separate local \hi{} from \hi{} associated with Phoenix. 

\subsection{Pegasus, Aquarius and Antlia}
Mixed type dIrr/dSph galaxies have generally been named as such because of the
presence of \hi{} centered on the optical disks of the galaxies. For that reason
these 3 galaxies do not hold many surprises.  \hi{} in the three galaxies was
previously reported by \citet[ for Pegasus and Aquarius]{lo93} and \citet[ for
Antlia]{barnes2001}. High resolution spectroscopy was not necessary for these
objects since the HIPASS detections were very clear and reliable.

However, Parkes narrowband observations have already been conducted on Antlia
and NGC 3109 by \citet{barnes2001b}. This galaxy is probably the most
interesting one of the 3 because of its proximity to NGC 3109.  Ionizing flux
coming from the larger companion may have disturbed the ISM of Antlia.
Unfortunately, the poor angular resolution and the lack of known optical
velocity for Antlia, make it very difficult to draw any conclusion on this
phenomenon apart from the fact that the \hi{} mass might be lower than what is
expected for that kind of galaxy.

\subsection{Fornax}
The Fornax galaxy is probably one of the most problematic in the LG, at least
for \hi{} studies. It has an optical systemic velocity of V$_{\odot} = 53\pm$3
\kms{}. In that direction and albeit at a Galactic latitude of $-66^{\circ}$,
the Milky Way's \hi{} radial velocity structure is quite extended and peaks at
V$_{\odot} \simeq 10$ \kms{}. The tail of this Galactic \hi{} line emission
extends beyond 53 \kms{} and is brighter than what would be expected from any
feature associated with the Fornax dwarf.

\begin{figure}[tbh]
\begin{center}
\includegraphics[angle=0, width=0.5\textwidth, height=0.5\textwidth]{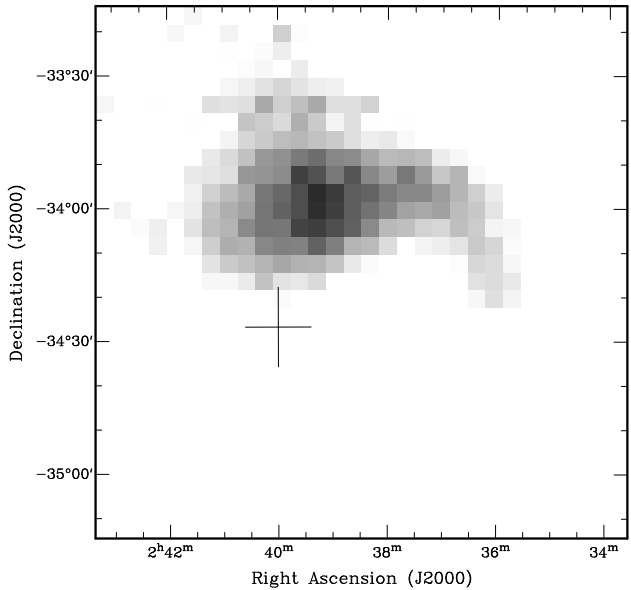}
\caption{
Velocity-integrated \hi{} line emission map of the Fornax dwarf region after subtraction of the MW
emission. The cross marks the position of the optical center
of the dwarf.}\label{fornaxframe}
\end{center}
\end{figure}

\begin{figure}[tbh]
\begin{center}
\includegraphics[angle=0, width=0.6\textwidth]{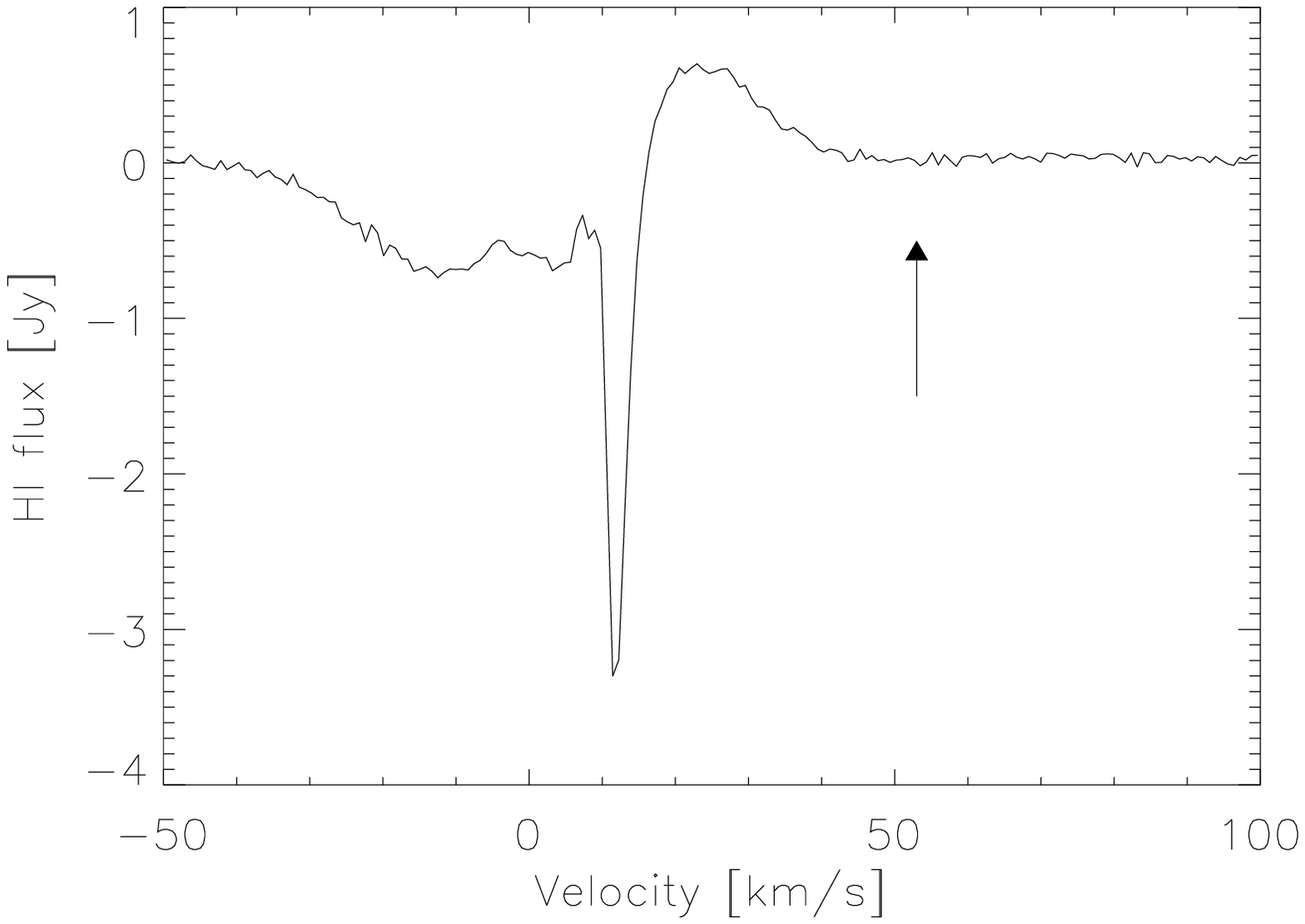}
\caption{The \hi{} spectra of the feature highlighted in Figure \ref{fornaxframe},
after removal of the MW emission. The subtraction left some residual near the 0
\kms{} velocity. The vertical arrow shows the optical velocity of Fornax.}\label{fornaxspectres}
\end{center}
\end{figure}

Figure \ref{fornaxframe} shows the integrated intensity map of Fornax after the
MW's contribution was removed. We see a twisted head-tail feature north of the
galaxy. The MW-subtracted spectrum in Figure \ref{fornaxspectres} shows that the
velocity of the excess \hi{} emission V$_{HI} = 23$ \kms{} is $\sim 30$ \kms{}
lower than the stellar value.

The 21 cm line shape of this cloud is obviously still contaminated by signatures
of the MW as the Galaxy's subtraction process is crude and the shape of the
interfering \hi{} emission badly determined. The best example of this is the
negative feature seen between -30 and 10 \kms{}. This is the result of a very
large \hi{} filament covering the whole south west quadrant of the cube that our
subtraction technique does not adequately remove.  Nevertheless, this technique
enables us to highlight this kinematically and structurally distinct cloud
projected well inside the 71$^{\prime}$ tidal radius \citep{irwin95} of the
Fornax dwarf. Interestingly, this cloud is closer to the center of the dwarf
than the extratidal stellar shell-like structure found by \citet{coleman2005}
1.3$^{\circ}$ northwest of the center. Apart from these apparent overlap, we are
unable to distinguish whether this cloud is: 1. part of the inherent clumpiness
of the MW, 2. part of the Fornax system or 3. a HVC somewhere in between.

If the \hi{} is indeed associated with Fornax, it has a \hi{} mass of
1.5$\times10^5~[d/138$ kpc$]^2$ \msol{}.  This should be regarded as a lower
limit because part of the mass resides below the 20 \kms{} mark where a very
strong MW feature is situated. The mass estimates, as well as the integrated
intensity map, have been calculated using data from above that velocity (and
below 40 \kms). If the \hi{} cloud is at the same radial distance as the Fornax
dwarf, the \hi{} mass, its position with respect to the stellar center ($\sim 1$
kpc) and the velocity difference between these two components are very similar
to what is found in other LG dwarf galaxies like Sculptor and Phoenix or the
Sculptor group dwarfs ESO294-G010 and ESO540-G030 \citep{bouchard2005}.

\subsection{Sextans and LeoI}
Sextans and LeoI were both reported as having gas associated with them (BR).
They argued that the velocity agreement and spatial overlap of the clouds with
the galaxies were sufficient evidences to prove that they were associated, the
chance of random agreement being around 1.5$\times10^{-3}$.  But careful
inspection of the HIPASS cube of the same regions reveals no trace of those
clouds. 

The Leiden-Dwingeloo survey with which the detections were initially made
has a sensitivity of 70 mK, much greater than HIPASS (rms noise of 10 mK).
The detections were of
low significance and were probably artefacts.

\section{Analysis}

\begin{table}[tb]
\caption{Properties of Local Group \hi{} clouds}\label{result}
\begin{minipage}{\linewidth}
\renewcommand{\thefootnote}{\thempfootnote}
\renewcommand{\footnoterule}{\rule{1cm}{0cm}\vspace{-0.2cm}}
\begin{tabular}{l c c c c c c}
\tableline
\tableline
Cloud		& L$_{V}$ & Distance	& $M_{\rm HI}$		& $V_{\odot}^{\rm HI}$& $P$~\tablenotemark{a}& HVC density\\%& Velocity Dispersion\\	
		& ($\times10^6$ L$_{\odot}$) & (kpc)		& ($\times10^6$ \msol)	& (\kms) & \% & (deg$^2$ \kms)$^{-1}$ \\
\tableline
Sculptor a	& 2.15	& 79$\pm$4	& 0.041$\pm$0.002 	& $100\pm$1	& 99.7				& 0.0022\\%& 4.0	\\
Sculptor b	& 2.15	& 79$\pm$4	& 0.193$\pm$0.003	& $105.1\pm$0.3	& 99.7				& 0.0022\\%& 6.2	\\
Cetus a	  	& 0.84	& 775$\pm$50	& 2.7$\pm$0.1		& $-311\pm$3	& \tablenotemark{\dagger}~72.1	& 0.0018\\%& 11.9	\\
Cetus b	  	& 0.84	& 775$\pm$50	& 0.74$\pm$0.08		& $-262\pm$9	& \tablenotemark{\dagger}~91.0	& 0.0018\\%& 7.8	\\
Cetus c	  	& 0.84	& 775$\pm$50	& 0.95$\pm$0.07		& $-268\pm$5	& \tablenotemark{\dagger}~44.5	& 0.0018\\%& 8.0	\\
Phoenix	  	& 0.90	& 445$\pm$30	& 0.19$\pm$0.01		& $-23\pm$1	& \nodata			& 0.0004\\%& \nodata	\\
Fornax	  	& 15.5	& 138$\pm$8	& 0.146$\pm$0.004 	& $29\pm$1	& \tablenotemark{\dagger}~87.7	& 0.0015\\%& 4.3 	\\
Tucana	  	& 0.55	& 880$\pm$40	& 1.8$\pm$0.1		& $132\pm$5	& \tablenotemark{\dagger}~98.7	& 0.0009\\%& 3.6 	\\
Pegasus	  	& 12.0	& 955$\pm$50	& 5.4$\pm$0.2		& $-186\pm$2	& \nodata			& \nodata\\%& 9.0 	\\	
LGS3 a	  	& 1.33	& 810$\pm$60	& 0.16$\pm$0.01		& $-288\pm$5	& \nodata			& \nodata\\%& 3.4 	\\
LGS3 b	  	& 1.33	& 810$\pm$60	& 0.50$\pm$0.02		& $-334\pm$1	& 99.1~\tablenotemark{b}	& \nodata\\%& 5.6	\\
LGS3 c	  	& 1.33	& 810$\pm$60	& 0.53$\pm$0.02		& $-361\pm$2	& 86.4~\tablenotemark{b}	& \nodata\\%& 5.4	\\
Aquarius  	& 0.81	& 800$\pm$250	& 1.8$\pm$0.2		& $-144\pm$3	& \nodata			& 0.0018\\%& 7.5 	\\
Carina a  	& 0.43	& 101$\pm$5	& 1.04$\pm$0.09		& $265\pm$6	& 31.5				& 0.0030\\%& 6.8	\\
Carina b  	& 0.43	& 101$\pm$5	& 0.09$\pm$0.02		& $256\pm$15	& 55.1				& 0.0030\\%& 7.4	\\
Carina c  	& 0.43	& 101$\pm$5	& 0.36$\pm$0.02		& $287\pm$4	& 0.6				& 0.0030\\%& 9.7	\\
Antlia	  	& 1.73	& 1235$\pm$65	& 0.62$\pm$0.03		& $362\pm$1	& \nodata			& 0.00005\\%& 3.6	\\
Sextans	  	& 0.50	& 86$\pm$4	& \nodata		& \nodata 		& \nodata		& 0.0007\\%& \nodata	\\
LeoI	  	& 4.79	& 250$\pm$30	& \nodata		& \nodata 		& \nodata		& \nodata\\%& \nodata	\\
\tableline
\tablenotetext{a}{Where values are noted with a dagger ($\dagger$),
the optical velocity information is not available. See text for details.}
\tablenotetext{b}{Instead of using a local probability as in the other cases, we
used the mean probability over the total extent of the HVC catalog because LGS3 
is situated outside of it.}
\end{tabular}
\end{minipage}
\end{table}

\subsection{\hi{} Content of \ds{} and \di/\ds}

It is not clear whether or not the detected \hi{} clouds in the line of sight of
dwarf galaxies are really associated with them. The mixed \di/\ds{} type is a
little easier. The \hi{} is found closer to their optical centers. The presence
(or lack of) \hi{} within a system is one of the main criteria used to
differentiate between these two types of dwarfs. Since many of these galaxies
show recent SF \citep{grebel98}, it should not be surprising to find substantial
amounts of \hi{} close to them.

BR had several non-detections. Among those, only one, Antlia, was on our target
list. It is now clear that Antlia has \hi{} associated with it. So the list of
non-detections in the LG, after being augmented by the two false detections made
by BR (i.e. Sextans and LeoI), consists of 9 galaxies. And I, And II, And VII,
Leo I, Leo II, Ursa Minor, Draco, Sagittarius and Sextans are now the only
members of the LG where no \hi{} has been found near their optical centers.
Among those, Leo I and Leo II are very interesting cases.  Despite the fact that
they are quite far from the Milky Way, (e.g. further away than Sculptor), no
\hi{} is found in their vicinity.  They are at comparable distances but have
different masses and luminosities. Whatever phenomenon was involved in the ISM
removal, it is likely to be linked mainly with environmental factors that would
be the same in both cases.  For And~I, And~II and And~VII, the proximity to
Andromeda is probably the major factor influencing the absence of \hi.
Similarly, Ursa Minor, Draco and Sextans are all at less than 100 kpc from the
Milky Way and no \hi{} was found near these galaxies. Finally, Sagittarius,
being in direct interaction with the Galaxy should have lost its \hi{} long
before it started loosing its stars.

\subsection{Tidal Interaction and Ram Pressure}

Different mechanisms have been proposed to remove gas from \ds{} galaxies.  The
most often talked about are tidal interactions with the more massive members of
the LG and ram pressure from the IGM.  

The \ds{} galaxies are often situated inside the gravitational potential of
a larger companion. This causes stars to escape the gravitational field of
the dwarf and are seen as extratidal features \citep[e.g. ][]{martinez2001,
palma2003}.  In that respect, Sculptor shows an interesting shape. It has
one cloud on one side and the other cloud on the other side. Because both
of them are aligned with the proper motion \citep{schweitzer95}, the shape
could be attributed to tidal stretching. However, optical photometry and
star counts show that the optical component of the galaxy is aligned in a
roughly perpendicular direction \citep{irwin95}.  Furthermore Sculptor does
not seem to have any extratidal features \citep{coleman2005b}. We therefore
believe that this phenomenon, by itself, does not play a significant role
in the gas removal from \ds. 

Otherwise, ram pressure could in principle be devastating for any \ds{} in
the vicinity of a large spiral. If its orbit leads a dwarf deep into the
halo (or even the disk) of a spiral, ram pressure will cause major gas-loss
in the dwarf.  In less extreme cases \citet{gallart2001} have estimated
that ram pressure will produce an \hi{} distribution that is offset from
the optical center, trailing behind the galaxy, relatively smooth and
slightly elongated in the direction of motion. 

The shape of the \hi{} distribution near Cetus, Tucana and Carina
correspond to these criteria up to the extent that we have no idea of the
direction and magnitude of their proper motion.  For Sculptor (cloud b) and
LGS3 (clouds b and c), the \hi{} clouds are elongated but perpendicular to
the direction to the optical center of the dwarfs.  For Fornax, if the
cloud was indeed associated with the dwarf and its position is the result
of ram pressure, the cloud must have been distorted by a change of
direction in the proper motion.

Strangely, amongst the above mentioned galaxies, Cetus, Tucana and LGS3 are
some of the most isolated galaxies in our sample and the most distant from
the MW. This is exactly the situation where ram pressure should normally be
minimised.  Moreover, Sculptor is the only \ds{} from our sample with a
known proper motion \citep{schweitzer95} and it appears that it might have
have a leading cloud as well as a trailing one.  It is difficult to explain
the presence of multiple \hi{} clouds using ram pressure and we conclude
that this mechanism alone cannot account for the \hi{} distribution that we
observe.

Ram pressure and tidal fields are not the only mechanisms one should
consider, internal mechanisms can also disrupt the \hi{} content of a
dwarf. Episodes of star formation, in particular if combined with
supernovae explosions can generate huge amounts of energy and remove the
\hi{} from the central regions of a dwarf. This will eventually make ram
pressure and tidal stripping more effective because the gas is expelled to
regions with even shallower potential well than the center of the dwarf.

\subsection{HVCs}

Some of the clouds presented in this paper are also listed in the High Velocity
Cloud (HVC) catalog of \citet{putman2002}.  This catalog lists 1997 \hi{} clouds
that do not fit simple Galactic rotation models. Of our 12 detected clouds that
lie within the limits of the HVC catalog (-500 \kms{} $< V_{LSR} <$ +500 \kms,
$|V_{LSR}|$ $>$ 90 \kms{} and $\delta <$ +2\deg) 8 are listed.  Not listed,
among the 40 galaxies that do have an entry, are the \di/\ds{} type galaxies
Pegasus and Antlia. Both Sculptor clouds are marked as potentially belonging to
the Sculptor \ds{} galaxy.

\begin{figure}[!t]
\begin{center}
\includegraphics[angle=0, width=0.5\textwidth]{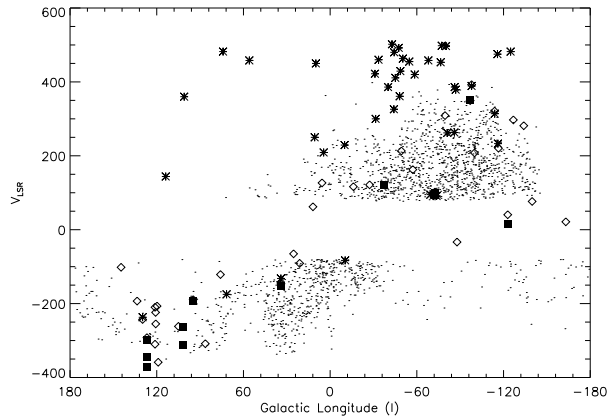}
\caption{The distribution of velocities of LG objects as a function of
Milky Way's Galactic Longitude. The dots represents the HVCs from the HVC
catalog by \citet{putman2002}. The asterisks are the galaxies listed in the same
catalog. Diamonds are LG galaxies \citep{mateo98} and solid squares are the
clouds listed in this paper that are believed to be associated with dwarf
galaxies.}\label{hvcdist}
\end{center}
\end{figure}

HVCs cover about 19\% of the southern sky and can be found in the same range of
velocities as the dwarf galaxies in the LG. Figure \ref{hvcdist} shows the
kinematic distribution of HVCs across the sky. The sinusoidal shape of the
diagram is caused by galactic rotation and the gap near V$_{LSR}$~=~0~\kms{}
comes from the limits imposed on the catalog.  In the Figure, the asterisks
represent galaxies that were listed in the HVC catalog. They have a distribution
that is systematically redshifted. Inspection of the catalog reveals that most
of them do not belong to the LG. The circles represents all the known LG
galaxies \citep{mateo98}. They have a different distribution then HVC/galaxies
but very similar to HVCs (dots) or our sample (triangles). 

For each galaxy we observed, we scanned the HIPASS HVC catalog and counted
the the number $N$ of HVCs near the target galaxies. We used these entries to
compute a local probability that none of these $N$ randomly distributed HVCs be
situated at the position or closer to the dwarf than the clouds we think might
be associated (these clouds are situated at $R_1$ radially and $\Delta V_1$
kinematically from the optical center of the galaxies).  It also assumes that
the HVCs are uncorrelated and distributed uniformly within the limits of our
search. Because we know that HVCs are not distributed uniformly across the sky
\citep[see ][]{putman2002}, we have limited our search to $R_0 < 10$\deg{} and
$\Delta V_0$ inside $V_{gal}~\pm~100$ \kms{} (or up to the catalog's limit)
around the galaxies in order to compute a local density of HVCs rather than
using an all-sky average.  The probability $P$ is given by:

\begin{equation}
\label{eq1}
P=\left[1-\left(\frac{R_1}{R_0}\right)^2 \frac{2\,\Delta V_1}{\Delta\! V_0}\right]^{N}
\end{equation}

\noindent The results are given in Table~\ref{result}. When the
optical velocity of the dwarf is unavailable, the value of $\frac{2\,\Delta
V_1}{\Delta\! V_0}$ is considered equal to 1 and no constraints on velocity is
applied when counting the clouds near the galaxy. Moreover, no $P$ is listed in
Table \ref{result} where no offset between the \hi{} and the optical could be
measured.  One should note that $P$ is computed for each of the listed cloud
independently, without the assumption that any other nearby clouds might be
associated (e.g. $P$ for Sculptor a does not assume that Sculptor b is
associated with the dwarf).

The galaxies where \hi{} clouds have been found $near$ them but not $in$ them
are: Sculptor, Cetus, Fornax, Tucana, LGS 3, and Carina. Apart for Carina, the
probabilities of these clouds genuinely belonging to the dwarfs are above 85\%.
For Cetus and Tucana, the probabilities do not take into account optical
velocity information because it is unavailable. Despite this fact, the
probability of association is very high and velocity information, if relatively
close to that of the cloud, would not drastically alter the probability.  We are
therefore confident that these clouds are associated with the corresponding
dwarf galaxies.

LGS 3 is situated outside of the boundaries of the catalog so the value listed
in Table~\ref{result} is based on an average HVC density. For LGS3 b, the
probability of association is higher than 99\% meaning that this association has
a good chance of being genuine. As we noted previously, the 2 clouds (LGS3 b and
c) seem to be related to each other. LGS3 c has a probability of association of
86.4\%. 

For Fornax, the velocity information was not used because the optical
velocity of the galaxy places it almost at the center of the velocity gap
of the catalog and the HVC count would therefore be unreliable. But even
without velocity information, the probability of association is 
very high (87.7\%) but not high enough to rule out the HVC hypothesis.

Cetus is one of the weak cases. Except for cloud b, the evidence is that
these clouds do not belong to the galaxy. Cloud b has a probability of
association of 91\% and seems to be ejected out of the dwarf in the
direction of cloud c. However, these probabilities have been computed
without any knowledge of radial velocity. If, as was stated in section 3.2,
the optical velocity of Cetus happens to be $-280\pm40$ \kms{} the
probabilities for all clouds could go up above 90\%. But, even if the optical velocity
does fall inside this bracket, further knowledge of the gas
removal mechanism will probably be required in order to undoubtedly confirm 
the cloud association.

The weakest case is that of Carina. Random HVC association will appear
in more than 40\% of the cases for any clouds. Chances are that Carina is
completely devoid of \hi. That has major consequences on the ability we
have to explain its SFH. \citet{hurley98} reported 3 distinct
episodes of star formation with a quiescent period of $\sim$4 Gyr. The last
of these episodes was around 2.5 to 3.5 Gyr ago. Yet there seems to be no
gas which could have fueled this SFH. The mechanisms by which the gas
could be removed must take these greatly disturbing facts into account.
\hi{} in Carina can disappear for a long period of time and then reappear to 
trigger SF. Carina b, if at the distance of Carina, would have a mass 
in the range expected for \ds{} but is situated far from it (1.8\deg,
projected distance of 3.2 kpc). Although it is not clear on how this gas
could fall back on the galaxy, it is quite easy to figure out how it could
have been expelled to that distance.

Simple kinematical studies cannot fully differentiate between the distribution
of galaxies and that of HVCs inside the LG. We therefore think that kinematical
arguments should be used with caution when discussing \ds{} and \hi{} clouds
association because of the systematic confusion that will emerge from such
correlation. It appears that dwarf galaxies from our sample are all situated in
regions where the HVC density is much higher (sometimes up to a factor of 30)
than that of the average density of HVC distribution ($\sim$0.0001 (deg$^{2}$
\kms)$^{-1}$). The only exception is Antlia that is a fairly isolated system.

The only obvious conclusion that can be drawn from Figure~\ref{hvcdist} and
Table~\ref{result} is that our cloud distribution is not different from that of
LG galaxies and therefore are not outside the LG or taking part in Galactic
rotation. To go further with this kind of kinematical study would require
knowledge on proper motions and models of galaxy formation that would include
accurate star formation models and stellar feedback.

\section{Conclusion}

Parkes observations were used to conduct a survey of the \ds{}
and \di/\ds{} galaxies of the Local Group. HIPASS and follow-up narrowband
observations enabled us to arrive at the following conclusions:

\begin{enumerate}

\item All LG \ds{} and \di/\ds{} of the southern hemisphere (except Antlia) are 
situated in phase-space regions where the HVC density is above average.

\item There are \hi{} clouds near (in projection) most of the \ds{} and 
\di/\ds{} of the Local Group. These clouds are significantly closer to
the galaxies than if they were randomly distributed HVCs. Although they are
often offset from the optical center they are probably physically
associated to the nearby galaxy.

\item The probability of the \hi{} being physically associated (and not being an
unrelated HVC projected near the dwarf) range from 99.7\% for Sculptor to less
than 1\% for Carina.  Overall, 11 out of the 16 investigated clouds have a
probability of being associated with the nearby galaxy that is above 85\%.

\item If the gas is associated with the galaxies, the offsets can not be
easily explained by neither ram pressure nor tidal stripping. Some of the
largest offsets are seen in galaxies where ram pressure should be the least
effective.

\end{enumerate}

\acknowledgments
We are grateful to the HIPASS team for giving us access to the survey data
files, to David Barnes for providing high resolution data on the Antlia dwarf
galaxy and to the referee for his helpful comments.

\newpage
\bibliographystyle{apj}

\begin{thebibliography}{47}
\expandafter\ifx\csname natexlab\endcsname\relax\def\natexlab#1{#1}\fi

\bibitem[{{Barnes} \& {de Blok}(2001)}]{barnes2001b}
{Barnes}, D.~G. \& {de Blok}, W.~J.~G. 2001, \aj, 122, 825

\bibitem[{{Barnes} {et~al.}(2001){Barnes}, {Staveley-Smith}, {de Blok},
  {Oosterloo}, {Stewart}, {Wright}, {Banks}, {Bhathal}, {Boyce}, {Calabretta},
  {Disney}, {Drinkwater}, {Ekers}, {Freeman}, {Gibson}, {Green}, {Haynes}, {te
  Lintel Hekkert}, {Henning}, {Jerjen}, {Juraszek}, {Kesteven}, {Kilborn},
  {Knezek}, {Koribalski}, {Kraan-Korteweg}, {Malin}, {Marquarding}, {Minchin},
  {Mould}, {Price}, {Putman}, {Ryder}, {Sadler}, {Schr{\"o}der}, {Stootman},
  {Webster}, {Wilson}, \& {Ye}}]{barnes2001}
{Barnes}, D.~G. {et al.} 2001, \mnras, 322, 486

\bibitem[{{Blitz} \& {Robishaw}(2000)}]{blitz2000}
{Blitz}, L. \& {Robishaw}, T. 2000, \apj, 541, 675

\bibitem[{{Bouchard} {et~al.}(2003){Bouchard}, {Carignan}, \&
  {Mashchenko}}]{bouchard2003}
{Bouchard}, A., {Carignan}, C., \& {Mashchenko}, S. 2003, \aj, 126, 1295

\bibitem[{{Bouchard} {et~al.}(2005){Bouchard}, {Jerjen}, {Da Costa}, \&
  {Ott}}]{bouchard2005}
{Bouchard}, A., {Jerjen}, H., {Da Costa}, G.~S., \& {Ott}, J. 2005, \aj, 130,
  2058

\bibitem[{{Br{\" u}ns} {et~al.}(2005){Br{\" u}ns}, {Kerp}, {Staveley-Smith},
  {Mebold}, {Putman}, {Haynes}, {Kalberla}, {Muller}, \&
  {Filipovic}}]{bruens2005}
{Br{\" u}ns}, {et al.} 2005, \aap, 432, 45

\bibitem[{{C{\^ o}t{\' e}} {et~al.}(2000){C{\^ o}t{\' e}}, {Mateo}, {Sargent},
  \& {Olszewski}}]{cote2000}
{C{\^ o}t{\' e}}, P., {Mateo}, M., {Sargent}, W.~L.~W., \& {Olszewski}, E.~W.
  2000, \apjl, 537, L91

\bibitem[{{Carignan} {et~al.}(1998){Carignan}, {Beaulieu}, {C{\^o}t{\'e}},
  {Demers}, \& {Mateo}}]{carignan98b}
{Carignan}, C., {Beaulieu}, S., {C{\^o}t{\'e}}, S.~., {Demers}, S., \& {Mateo},
  M. 1998, \aj, 116, 1690

\bibitem[{{Carignan} {et~al.}(1991){Carignan}, {Demers}, \&
  {Cote}}]{carignan91}
{Carignan}, C., {Demers}, S., \& {Cote}, S. 1991, \apjl, 381, L13

\bibitem[{{Coleman} {et~al.}(2005){Coleman}, {Da Costa}, {Bland-Hawthorn}, \&
  {Freeman}}]{coleman2005}
{Coleman}, M.~G., {Da Costa}, G.~S., {Bland-Hawthorn}, J., \& {Freeman}, K.~C.
  2005, \aj, 129, 1443

\bibitem[{{Coleman} {et~al.}(2005){Coleman}, {Da Costa}, \&
  {Bland-Hawthorn}}]{coleman2005b}
{Coleman}, M.~G., {Da Costa}, G.~S., \& {Bland-Hawthorn}, J. 2005, \aj, 130,
  1065

\bibitem[{{D'Ercole} \& {Brighenti}(1999)}]{dercole99}
{D'Ercole}, A. \& {Brighenti}, F. 1999, \mnras, 309, 941

\bibitem[{{Gallart} {et~al.}(2001){Gallart}, {Mart\'\i{}nez-Delgado}, {G{\'
  o}mez-Flechoso}, \& {Mateo}}]{gallart2001}
{Gallart}, C., {Mart\'\i{}nez-Delgado}, D., {G{\' o}mez-Flechoso}, M.~A., \&
  {Mateo}, M. 2001, \aj, 121, 2572

\bibitem[{{Grebel}(1998)}]{grebel98}
{Grebel}, E.~K. 1998, Highlights in Astronomy, 11, 125

\bibitem[{{Grebel}(2001)}]{grebel2001}
---. 2001, Astrophysics and Space Science Supplement, 277, 231

\bibitem[{{Harbeck} {et~al.}(2004){Harbeck}, {Gallagher}, \&
  {Grebel}}]{harbeck2004}
{Harbeck}, D., {Gallagher}, J.~S., \& {Grebel}, E.~K. 2004, \aj, 127, 2711

\bibitem[{{Hensler} {et~al.}(2004){Hensler}, {Theis}, \&
  {Gallagher}}]{hensler2004}
{Hensler}, G., {Theis}, C., \& {Gallagher}, J.~S. 2004, \aap, 426, 25

\bibitem[{{Hulsbosch} \& {Wakker}(1988)}]{hulsbosch88}
{Hulsbosch}, A.~N.~M. \& {Wakker}, B.~P. 1988, \aaps, 75, 191

\bibitem[{{Hurley-Keller} {et~al.}(1999){Hurley-Keller}, {Mateo}, \&
  {Grebel}}]{hurley99}
{Hurley-Keller}, D., {Mateo}, M., \& {Grebel}, E.~K. 1999, \apjl, 523, L25

\bibitem[{{Hurley-Keller} {et~al.}(1998){Hurley-Keller}, {Mateo}, \&
  {Nemec}}]{hurley98}
{Hurley-Keller}, D., {Mateo}, M., \& {Nemec}, J. 1998, \aj, 115, 1840

\bibitem[{{Irwin} \& {Hatzidimitriou}(1995)}]{irwin95}
{Irwin}, M. \& {Hatzidimitriou}, D. 1995, \mnras, 277, 1354

\bibitem[{{Johnston} {et~al.}(1999){Johnston}, {Sigurdsson}, \&
  {Hernquist}}]{johnston99}
{Johnston}, K.~V., {Sigurdsson}, S., \& {Hernquist}, L. 1999, \mnras, 302, 771

\bibitem[{{Larson}(1974)}]{larson74}
{Larson}, R.~B. 1974, \mnras, 169, 229

\bibitem[{{Lo} {et~al.}(1993){Lo}, {Sargent}, \& {Young}}]{lo93}
{Lo}, K.~Y., {Sargent}, W.~L.~W., \& {Young}, K. 1993, \aj, 106, 507

\bibitem[{{Majewski} {et~al.}(2000){Majewski}, {Ostheimer}, {Patterson},
  {Kunkel}, {Johnston}, \& {Geisler}}]{majewski2000}
{Majewski}, S.~R., {Ostheimer}, J.~C., {Patterson}, R.~J., {Kunkel}, W.~E.,
  {Johnston}, K.~V., \& {Geisler}, D. 2000, \aj, 119, 760

\bibitem[{{Mart\'\i{}nez-Delgado} {et~al.}(1999){Mart\'\i{}nez-Delgado}, {Gallart}, \&
  {Aparicio}}]{martinez99}
{Mart\'\i{}nez-Delgado}, D., {Gallart}, C., \& {Aparicio}, A. 1999, \aj, 118, 862

\bibitem[{{Mart{\'{\i}}nez-Delgado} {et~al.}(2001){Mart{\'{\i}}nez-Delgado},
  {Alonso-Garc{\'{\i}}a}, {Aparicio}, \& {G{\'o}mez-Flechoso}}]{martinez2001}
{Mart{\'{\i}}nez-Delgado}, D., {Alonso-Garc{\'{\i}}a}, J., {Aparicio}, A., \&
  {G{\'o}mez-Flechoso}, M.~A. 2001, \apjl, 549, L63

\bibitem[{{Mateo} {et~al.}(1998){Mateo}, {Olszewski}, \& {Morrison}}]{mateo98b}
{Mateo}, M., {Olszewski}, E.~W., \& {Morrison}, H.~L. 1998, \apjl, 508, L55

\bibitem[{{Mateo}(1998)}]{mateo98}
{Mateo}, M.~L. 1998, \araa, 36, 435

\bibitem[{{Matzner} \& {McKee}(2000)}]{matzner2000}
{Matzner}, C.~D. \& {McKee}, C.~F. 2000, \apj, 545, 364

\bibitem[{{Moore} {et~al.}(1999){Moore}, {Ghigna}, {Governato}, {Lake},
  {Quinn}, {Stadel}, \& {Tozzi}}]{moore99}
{Moore}, B., {Ghigna}, S., {Governato}, F., {Lake}, G., {Quinn}, T., {Stadel},
  J., \& {Tozzi}, P. 1999, \apjl, 524, L19

\bibitem[{{Mould} {et~al.}(1990){Mould}, {Bothun}, {Hall}, {Staveley-Smith}, \&
  {Wright}}]{mould90}
{Mould}, J.~R., {Bothun}, G.~D., {Hall}, P.~J., {Staveley-Smith}, L., \&
  {Wright}, A.~E. 1990, \apjl, 362, L55

\bibitem[{{Murali}(2000)}]{murali2000}
{Murali}, C. 2000, \apjl, 529, L81

\bibitem[{{Navarro} {et~al.}(1996){Navarro}, {Frenk}, \& {White}}]{navarro96}
{Navarro}, J.~F., {Frenk}, C.~S., \& {White}, S.~D.~M. 1996, \apj, 462, 563+

\bibitem[{{Oosterloo} {et~al.}(1996){Oosterloo}, {Da Costa}, \&
  {Staveley-Smith}}]{oosterloo96}
{Oosterloo}, T., {Da Costa}, G.~S., \& {Staveley-Smith}, L. 1996, \aj, 112,
  1969

\bibitem[{{Palma} {et~al.}(2003){Palma}, {Majewski}, {Siegel}, {Patterson},
  {Ostheimer}, \& {Link}}]{palma2003}
{Palma}, C., {Majewski}, S.~R., {Siegel}, M.~H., {Patterson}, R.~J.,
  {Ostheimer}, J.~C., \& {Link}, R. 2003, \aj, 125, 1352

\bibitem[{{Putman} {et~al.}(2002){Putman}, {de Heij}, {Staveley-Smith},
  {Braun}, {Freeman}, {Gibson}, {Burton}, {Barnes}, {Banks}, {Bhathal}, {de
  Blok}, {Boyce}, {Disney}, {Drinkwater}, {Ekers}, {Henning}, {Jerjen},
  {Kilborn}, {Knezek}, {Koribalski}, {Malin}, {Marquarding}, {Minchin},
  {Mould}, {Oosterloo}, {Price}, {Ryder}, {Sadler}, {Stewart}, {Stootman},
  {Webster}, \& {Wright}}]{putman2002}
{Putman}, M.~E. {et al.} 2002, \aj, 123, 873

\bibitem[{{Sarajedini} {et~al.}(2002){Sarajedini}, {Grebel}, {Dolphin},
  {Seitzer}, {Geisler}, {Guhathakurta}, {Hodge}, {Karachentsev},
  {Karachentseva}, \& {Sharina}}]{sarajedini2002}
{Sarajedini}, A., {et al.} 2002, \apj, 567, 915

\bibitem[{{Schweitzer} \& {Cudworth}(1996)}]{schweitzer96}
{Schweitzer}, A.~E. \& {Cudworth}, K.~M. 1996, in American Astronomical Society
  Meeting, Vol. 188, 0901

\bibitem[{{Schweitzer} {et~al.}(1997){Schweitzer}, {Cudworth}, \&
  {Majewski}}]{schweitzer97}
{Schweitzer}, A.~E., {Cudworth}, K.~M., \& {Majewski}, S.~R. 1997, in ASP Conf.
  Ser. 127: Proper Motions and Galactic Astronomy, 103

\bibitem[{{Schweitzer} {et~al.}(1995){Schweitzer}, {Cudworth}, {Majewski}, \&
  {Suntzeff}}]{schweitzer95}
{Schweitzer}, A.~E., {Cudworth}, K.~M., {Majewski}, S.~R., \& {Suntzeff}, N.~B.
  1995, \aj, 110, 2747

\bibitem[{{St-Germain} {et~al.}(1999){St-Germain}, {Carignan}, {C{\^o}te}, \&
  {Oosterloo}}]{julie99}
{St-Germain}, J., {Carignan}, C., {C{\^o}te}, S.~., \& {Oosterloo}, T. 1999,
  \aj, 118, 1235

\bibitem[{{Staveley-Smith} \& {et al.}(2000)}]{staveley2000}
{Staveley-Smith}, L. \& {et al.} 2000, in ASP Conf. Ser. 217: Imaging at Radio
  through Submillimeter Wavelengths, 50

\bibitem[{{Stetson}(1980)}]{stetson80}
{Stetson}, P.~B. 1980, \aj, 85, 387

\bibitem[{{Tolstoy} \& {Irwin}(2000)}]{tolstoy2000}
{Tolstoy}, E. \& {Irwin}, M. 2000, \mnras, 318, 1241

\bibitem[{{Whiting} {et~al.}(1999){Whiting}, {Hau}, \& {Irwin}}]{whiting99}
{Whiting}, A.~B., {Hau}, G. K.~T., \& {Irwin}, M. 1999, \aj, 118, 2767

\bibitem[{{Young} \& {Lo}(1997)}]{young97a}
{Young}, L.~M. \& {Lo}, K.~Y. 1997, \apj, 476, 127

\bibitem[{{Zaritsky}(1999)}]{zaritsky99}
{Zaritsky}, D. 1999, in ASP Conf. Ser. 165: The Third Stromlo Symposium: The
  Galactic Halo, 34

\end{thebibliography}

\end{document}